\documentstyle[aasms4,flushrt]{article}

\lefthead{MOOKERJEA \& GHOSH}
\righthead{S.E.D. of IRAS 18314--0720, 18355--0532 \& 18316--0602}

\begin{document}

\title{Understanding the Spectral Energy Distributions of the Galactic Star 
Forming Regions IRAS 18314--0720, 18355--0532 \& 18316--0602}

\author{Bhaswati Mookerjea \& S.K. Ghosh}
\centerline{\it Tata Institute of Fundamental Research}
\centerline{\it  Dr. Homi Bhabha Road, Colaba}
\centerline{\it Mumbai (Bombay) 400 005, India}

\begin{abstract}
 Embedded Young Stellar Objects (YSO) in dense interstellar 
clouds is treated self-consistently to understand their
spectral energy distributions (SED). Radiative transfer 
calculations in spherical geometry involving the dust as 
well as the gas component, have been carried out to 
explain observations covering a wide spectral range 
encompassing near-infrared to radio continuum wavelengths. 
Various geometric and physical details of the YSOs are
determined from this modelling scheme.

  In order to assess the effectiveness of this
self-consistent scheme, three young
Galactic star forming regions associated with 
IRAS 18314--0720, 18355--0532 and 18316--0602 have been modelled as
test cases. They cover a large range of luminosity
($\approx$ 40). 
The modelling of their SEDs has
led to information about various details of these
sources, e.g. embedded energy source, cloud
structure \& size, density distribution,
composition \& abundance of dust grains etc.
In all three cases, the best fit model corresponds to
the uniform density distribution. 
Two types of dust have been considered, viz., Draine \&
Lee (DL) and the Mezger, Mathis \& Panagia (MMP).
Models with MMP type dust explain the dust continuum
and radio continuum emission from IRAS 18314--0720 \&
18355--0532 self-consistently.
These models predict much lower intensities for the
fine structure lines of ionized heavy elements,
than those observed for IRAS 18314--0720 \& 18355--0532.
This discrepancy has been resolved by invoking 
clumpiness in the interstellar medium.
For IRAS 18316--0602, the model with DL type dust
grains is preferred.

\end{abstract}

{\em Subject Headings}: {Infrared SED -- H II regions -- Radiative Transfer
-- IRAS 18314--0720 -- IRAS 18355--0532 -- IRAS 18316--0602}

\section{Introduction}

   The formation and the initial stages of evolution of
stars take place inside dense regions within molecular 
clouds, from which they are born. Hence, the star formation
studies have necessarily to deal with the protostars/young
stars in the environment of interstellar gas and dust
from the parent cloud. The stars with sufficient supply of
Lyman continuum photons (basically depending on their mass)
create and maintain H II regions around them.
The (ultra)/compact variety of H II regions in particular
provides a natural test case for
better understanding of medium / high mass star formation.
They present two main advantages for such studies: being
younger they allow one to probe the physical processes closer in
time to the formation of the embedded star, and secondly they have
simpler geometrical shapes (e.g. with spherical/cylindrical 
symmetry), thus providing opportunity for direct comparison 
between the observations and
the predictions of detailed numerical models. In addition,
being on the higher luminosity class (roughly between ZAMS O4 to 
B0.5), observationally they can be studied over larger distances 
(almost any part of the Galaxy) in the radio continuum as well as 
infrared/sub-mm.
  Available observational data regarding 
continuum emission from Galactic compact H II regions
span a wide spectral range covering near-infrared (NIR) to 
sub-millimeter wavelengths. Whereas the atmospheric windows in 
NIR, sub-millimeter and a few restricted ones in the mid-IR (MIR)
have been used extensively, the peak of the spectral energy
distribution (SED) lies invariably
in the far-IR (FIR) waveband which is 
inaccessible from the earth based observatories. The work horse
for the high angular resolution FIR measurements of Galactic 
star forming regions 
has been the Kuiper Airborne Observatory,
occasionally supplemented by balloon borne telescopes. Although
the InfraRed Astronomy Satellite (IRAS)
offers the only near all sky and most complete data base,
its poorer angular resolution limits its usefulness, particularly
for the study of compact H II regions. However, this situation 
is about to change drastically 
with the advent of the European Space Agency's Infrared 
Space Observatory (ISO) mission.
ISO with improved photometric sensitivity as well as spectral and angular
resolution, will open up enormous opportunity to perform 
detailed studies of Galactic star forming regions in the 
entire infrared waveband encompassing Near through Mid to Far Infrared. 
 The most of the infrared continuum emission from the
embedded YSOs originating from the interstellar dust
component will be measured throughout the
2.5 -- 200 $\mu$m region with greater photometric 
precision  and with diffraction limited angular 
resolution, by the ISO instruments. In addition,
the spectroscopic data will help understand
the composition as well as physical parameters of the
interstellar gas component.

  Keeping the above in mind, the present study is a step in 
trying to establish a self-consistent yet simple scheme of modelling
Galactic compact H II regions in spherically symmetric geometry.
Here, by self-consistent we mean that the same geometric
and physical configuration fits the  observed
data for the emission from the dust (most of the
infrared, sub-mm, mm part of the SED) as well as the
emission from the gas (radio continuum, fine structure
line etc).
The main aim is to extract maximum possible information
regarding the geometric and physical details of the
star forming region from the observed SED.
In addition to 
predicting the continuum infrared
emission from the dust, the gas component has also been 
integrated in a self-consistent manner,
thereby predicting the absolute and relative strengths
of atomic/nebular lines as well as the radio continuum emission.

   In order to assess the usefulness of the above scheme,
three Galactic embedded YSOs (IRAS 18314--0720, 18355--0532
\& 18316--0602; most likely in compact
H II region phase, have been selected for modelling. These
sources have been chosen based on availability of relevant
observational data; and also to cover a wide range of total
luminosity (a factor of $\approx$ 40). 
These three sources IRAS 18314--0720, 18355--0532 \& 18316--0602
have luminosities 1.02 $\times 10^{6}$, 1.4 $\times 10^{5}$
\& 2.5 $\times 10^{4} L_{\odot}$ respectively 
(which correspond to single ZAMS stars of type
O4, O6.5 and B0).
Although at this point of time the available
amount of validated ISO data in public domain is rather limited,
much more sophisticated application
of this scheme is anticipated in the future.

 The outline of this paper is as follows.
 The section 2 describes the modelling scheme including
the treatments of radiative
transfer in the dust and the gas components respectively. 
The section 3 presents the observational constraints
of the three sources,
and the results of modelling in the form of
geometrical and physical
information extracted about these sources.
Conclusions are summarized in section 4.


\section{The Modelling Scheme}

A compact H II region is modelled as a
spherically symmetric cloud (made of typical interstellar gas-dust
material), powered by centrally embedded single or a cluster
of zero age 
main sequence (ZAMS) star/(s). This cloud is assumed to be immersed
in an isotropic radiation field (typical Interstellar Radiation
Field, ISRF). The interstellar gas and the dust is assumed
to follow the same radial density distribution law, but
with the following difference $\--$
whereas the gas exists throughout the cloud (i.e. right from the
stellar surface upto the outer boundary of the cloud, $R_{max}$;
see Figure 1), there is a natural lower limit to the inner 
boundary, $R_{min}$, for
the dust distribution (i.e. a cavity in the dust cloud).
This is because the dust grains
are destroyed when exposed to excessive radiative heating.
The gas to dust ratio, where they co-exist ($ R_{min} < r <
R_{max}$), is assumed to be constant. The position of the
ionization front ($R_{HII}$, refer to Figure 1) depends on
the effective temperature and luminosity of the exciting star, 
as well as the density of the gas. The case,
$R_{HII}$ $<$ $R_{min}$ is also possible, if either the star 
is not hot enough and/or
the density of gas around the star is quite high. 

Modelling a specific compact H II region involves 
matching the predicted emergent spectral (continuum) shape 
with the measured SED
and comparing the relative and absolute strengths of the 
atomic / nebular spectral lines from the gas component
with spectroscopic observations (if available).

   The following are the inputs for individual runs of
the modelling scheme :
(i) the total luminosity
(ii) the spectral shape of the radiation
emerging from the embedded source/(s)
(sensitive to the assumed Initial Mass Function and
upper mass cut off, in case of a star cluster);
(iii) the gas to dust ratio;
(iv) the properties of the dust grains (of each type considered);
(v) the ISRF incident at the outer boundary; 
(vi) the elemental abundances in the gas component (which is assumed 
to be uniform throughout the cloud).

 The total luminosity of the embedded energy source/(s)
is frozen at the value determined by integrating the
observed continuum SED.
 The embedded energy source/(s) is varied between a single
ZAMS star and a cluster of ZAMS stars. The canonical value
of 100~:~1 for the gas to dust ratio by mass is used initially, but
varied if no acceptable model can be constructed to fit all
the data.
 Two types of dust have been considered here, that of
Draine \& Lee (1984; hereafter DL) and Mezger, Mathis \& Panagia
(1982; hereafter MMP). Within each type, dust consists of
grains of different composition and size (see section 2.2.1).
 The ISRF has been taken from Mathis, Mezger \& Panagia (1983),
and has been held fixed for all model runs. 
The gas component is assumed to be
consisting of either (a) only
hydrogen (section 2.2.1); or (b) typical H II region abundance 
as listed by Ferland (1996) (section 2.2.2). 
In the latter case, only elements
with abundance (relative to hydrogen) higher than 
$3.0 \times {10}^{-6}$ have been considered.
In some model runs for a specific H II region, 
the elemental abundances used by earlier worker/(s) have been used
for comparison.
Although the elemental abundance is same throughout the cloud,
the ionization structure and the various level
populations depend on several physical parameters
including the local radiation field.

The following parameters are explored in order
to get an acceptable fit to all the data :
(i) geometric details like $R_{max}$ \& $R_{min}$ ($R_{min}$
will not violate radiative destruction of grains);
(ii) radial density distribution law (only three power laws
have been explored, viz., $n(r) \approx r^{0}, r^{-1}$ or
$r^{-2}$);
(iii) total radial optical depth due to the dust (inclusive of all
constituents) 
at any selected wavelength; \&
(iv) the dust composition or relative fractions of
different constituent grain types (see section 2.1.1).

The interstellar cloud is divided into 141 radial grid points.
Near both the boundaries, these grid points are
logarithmically spaced (in rest of the cloud, a linear grid has
been used). The frequency grid consists of 89 points
covering the wavelength range 944 \AA ~ to 5000 $\mu$m.

\subsection{ Radiative transport through the dust component (D1)}
  The radiative transport through the dust component
has been carried out by using a programme based on the code CSDUST3 
(Egan, Leung \& Spagna 1988). We have improvised this code 
by generalizing the boundary conditions leading to 
much better flexibility for modelling typical
astrophysical sources.
The moment equation of radiation transport and the equation
of energy balance are solved simultaneously as a two-point 
boundary value problem in this programme.
The effects
of multiple scattering, absorption and re-emission of
photons on the temperature of dust grains and the internal
radiation field have been considered.
In addition, the following details, viz., the radiation 
field anisotropy, linear anisotropic 
scattering and multi grain components are also included.
The entire relevant spectral range covering right from
the UV wavelengths to the millimeter region has
been considered (the frequency grid consists of 89 points).

For preserving the energetics precisely and self-consistently, 
the total energy 
available for heating of the dust component includes
all three components (all components being binned into
the respectively relevant spectral grid elements) :
(i) the star cluster / ZAMS stellar luminosity in photons 
below the Lyman limit ($\lambda > 912 $\AA);
(ii) a part of the Lyman continuum luminosity of the embedded star, 
($\lambda <912 $\AA), directly absorbed by the dust; and
(iii) a fraction of the same reprocessed by the gas.
The last contribution, viz., the reprocessed Lyman continuum 
photons, has been quantified by the prescription of
Aller \& Liller (1969) that each Lyman 
continuum photon emitted by
the star ultimately leads to one Ly-$\alpha$ photon 
and one Balmer-$\alpha $ photon.\\

  From the resulting dust temperature distribution in the
cloud, the emergent intensities as a function of frequency
at various impact parameters (depending on the radial grid)
are calculated.
Hereafter, the above modelling scheme is referred to as ``D1".

\subsubsection{Dust Grains}
 Two different approaches have been taken to deal with
the dust grains in the present study.
The first approach, referred to as ``DL" (since largely
based on grain properties of Draine \& Lee (1984)), 
is summarized below.  The physical 
properties of the grains, viz., absorption and scattering 
efficiencies, $Q_{abs}(a, \nu), Q_{sca}(a, \nu)$, 
and the scattering anisotropy factor,
$g(a, \nu)$, for all sizes ($a$) and
frequencies ($\nu$) have been taken from the 
tables of B.T. Draine's home page which are computed
similar to Laor \& Draine (1993). These have a finer 
grid of grain sizes than Draine \& Lee (1984).
  Three types of most commonly accepted variety of
interstellar dust have been included in this DL
case, viz.,
(i) Graphite, (ii) Astronomical Silicate \& (iii)
Silicon Carbide (SiC). The relative abundances of
these three types of grains are used as parameters
to fit the observed SED well.

   In the second approach, the dust grain properties
proposed by Mezger, Mathis \& Panagia (1982) have been used.
This approach, hereafter ``MMP", has been considered
since it has been found very useful
in explaining the SED of certain class of YSO's
(Butner et al 1991). This type of dust consists
of graphite and silicate only, but their absorptive
and scattering properties differ substantially 
from those for the DL case.

The size distribution 
of the dust grains is assumed in accordance with 
Mathis, Rumpl \& Nordsiek (1977), to be a power law, viz.,
$n(a)da \sim a^{-m}da$, $a_{min} \leq a \leq a_{max}$
with $m$ = 3.5. The lower and upper limits of the
grain size distribution $a_{min}$ and $a_{max}$
have been chosen as recommended by 
Mathis, Mezger \& Panagia (1983), 
to be 0.01 $\mu m$ and 0.25 $\mu m$ respectively. 


\subsection{Radiative transport through the gas component}

  Two independent approaches have been taken to deal with 
the radiative transfer through the gas component in spherical geometry. 
The first one takes a very simplistic view considering
photoionization and recombination of hydrogen alone, neglecting
other heavier elements,
as well as the gas-dust coupling (where they co-exist). 
The treatment of this first approach has been
entirely developed in the present study. 

The second approach is more sophisticated and considers
several prominent elements in the gas phase of the cloud.
In addition to photoionization and recombination,
other physical processes like collisional excitation \& 
de-excitation, grain photoionization and gas-dust coupling
are also included.
This detailed modelling involves the
use of the photoionization code CLOUDY (Ferland, 1996),
which has been supplemented with a software scheme developed 
in the present work, to make the model predictions more
realistic and easy to compare with observations.

 Whereas the first approach has been used to compute the
expected radio continuum emission (without any
assumption about optical thinness of the gas at radio
wavelength), the second one is used for predicting
the atomic / ionic nebular line emission strengths.
Both these approaches are discussed next.

\subsubsection{The simple approach (G1)}
In the simple approach, since only hydrogen has been
considered, the ionization structure of the gas can
be specified by, $R_{HII}$, the location of the
boundary of the H II region (see Figure 1). 
$R_{H II}$ has been determined by
considering the radiative transfer of Lyman continuum photons
from the embedded star cluster / ZAMS star through the cloud. 
The effect of the dust component in extinguishing the 
radiation field, 
has been considered whenever necessary (cases satisfying 
$R_{min} < R_{HII} < R_{max})$. 
In addition, the radio continuum emission has been computed
including the effect of appropriate radio optical depth (self
absorption). Emission of relevant
recombination lines from the ionized gas has been 
quantified for their role in the radiative transfer 
through the dust component.
Hereafter, this simplistic modelling of the radiative transfer
of UV and radio through the interstellar cloud
will be referred to as ``G1".
Appendix 1 gives more details of this treatment.

\subsubsection{The detailed approach (G2)}
  For the detailed approach to the radiative transfer
through the gas component,
the code CLOUDY (Ferland 1996), supplemented by a software
scheme developed in the present study, has been used. 
The supplementary scheme improves the modelling 
by (i) emulating the exact structure of the compact H II
region; and (ii) including
 the absorption effects of the dust (present within the 
line emitting zones), on the emergent line intensities.
This detailed approach 
self-consistently deals with almost all physical processes
(radiative - collisional equilibrium) important 
in and around a photoionized nebula. It simultaneously
looks for statistical and thermal equilibrium by solving
the equations balancing ionization-neutralization
processes and heating-cooling processes. It predicts
physical conditions of the gas, e.g. ionization, 
level populations,
temperature structure, and the emerging emission line
spectrum.
     The gas component of the cloud has been 
considered with typical H II
region abundance, as tabulated in Ferland (1996). 
This is an average of Baldwin et al. 1991, 
Osterbrock, Tran \& Veilluex 1992,
and Rubin et al. 1991, unless specified otherwise. 
Only the elements with abundance 
relative to hydrogen,  higher than $3.0 \times 10^{-6}$ have been
used. This results into the following elements : H, He,
C, N, O, Ne, Mg, Si, S and Ar. The grains of the 
Astronomical Silicate
and Graphite types have been introduced at and beyond 
a radial distance from the exciting star such that they
do not heat up above their sublimation temperature.
The heating (photoelectric) and cooling (collisional)
due to grains have been considered. The effect of a
constant cosmic ray density on the gas is included
(which affects energy deposition and ionization).

To be self-consistent with the radiative transfer
treatment through the dust component (D1),
the entire cloud is considered to be consisting of
two spherical shells, the inner one made of gas alone and
the outer one with gas and dust. The boundary between
the two shells,
$R_{min}$, is taken from the corresponding 
best fit D1(DL) or D1(MMP) model.
CLOUDY
is run twice, the first time (RUN1) for the inner pure gas
shell with the central energy source. 
The continuum emerging from RUN1 is used as input to
the second run (RUN2) for the outer shell.
The emerging line spectrum from RUN1 is transported 
to an outside observer,
through the second (outer) shell by considering the 
extinction due to the entire dust column present there.
For every spectral line considered, its emissivities 
from individual radial zones of RUN2
are transported through the corresponding remaining 
dust column densities within the outer shell.
The emerging line luminosities from RUN1 and RUN2 are
finally added to predict the total observable luminosity.
This detailed modelling scheme will be referred to
as ``G2" in the later text.

A total of 27 most prominent spectral lines (from various
ionization stages of the above mentioned 10 elements) 
have been considered. From observational point of view, 
the reliable detectability of any spectral line will
depend on experimental detail like : 
the instrument line function (spectral resolution);
as well as the strength of the continuum
in the immediate spectral neighbourhood of the line. 
An attempt has been made to predict line intensities
for those lines which are detectable by
the spectrometers onboard ISO (SWS \& LWS).

A line has been defined to be ``detectable" only if the
expected power incident on the detector, due to the spectral
line is more than 1\% of the continuum (from the same
astrophysical source, originating from the dust) 
in the corresponding resolution element. 
%
%
An instrumental resolution in the range 1000 -- 20000 has been 
reported for ISO spectrometers
depending on the wavelength (de Graauw et al. 1996; Swinyard et al.
1996).
The lines are in general narrower than
the resolution element if the widths are thermal.
Only the lines turning out to be ``detectable" according 
to the above criteria are presented with details.


\section{Study of the Galactic Star Forming regions :
IRAS 18314--0720, 18355--0532 \& 18316--0602}

 With the aim of extracting important geometrical and
physical details of the Galactic star forming regions --
IRAS 18314--0720, 18355--0532 \& 18316--0602, the 
modelling scheme described earlier, has been applied.
These sources have been selected to cover a range
of $\approx$ 40 in the total luminosity. In addition, they
have adequate observational data necessary to 
constrain the modelling. In what follows, the observations
available for these sources and the results of modelling
them, are described.

\subsection{IRAS 18314--0720}

 The IRAS Point Source Catalog (hereafter, IRAS PSC) source
18314--0720 has flux densities of 156, 648, 4714 \& 8089 Jansky
in 12, 25, 60 \& 100 $\mu$m bands respectively. Although this 
source did not appear in the original IRAS Low Resolution
Spectra Atlas (8 -- 22 $\mu$m; hereafter LRS), later analysis
released the LRS spectrum of this source (Volk \& Cohen, 1989). 
Based on LRS spectrum, forbidden lines of ions of neon and sulphur 
have been identified as
well as possible detection of features due to
Polycyclic Aromatic Hydrocarbons have been reported
(Jourdain de Muizon, Cox \& Lequeux 1990). 
Recently, mid \& far infrared spectroscopic 
detection of several ionic lines based on Kuiper
Airborne Observatory measurements have become 
available (Afflerbach, Churchwell \& Werner, 1997).
IRAS 18314--0720 corresponds to the Revised Air Force
Geophysical Laboratory  source RAFGL 2190, which was
detected in 4.2, 11, 20 \& 27 $\mu$m bands.
IRAS 18314--0720 was included in the IRAS colour selected
sample of Chini et al (1986) for study at 1.3 mm continuum
and near infrared mapping (Chini, Krugel \& Wargau 1987).
Bronfman, Nyman \& May (1996) have detected $CS$ emission 
at 98 GHz from IRAS 18314--0720.
In a search for $NH_{3}$ and $H_{2}O$ maser sources associated
with this source, the former has been detected but not
the latter (Churchwell, Walmsley \& Cesaroni 1990).

The radio continuum emission from the H II region associated with
IRAS 18314--0720 has been observed at various frequencies.
The IRAS PSC associates 18314--0720 with radio continuum sources 
of Parkes and Bonn
surveys of the Galactic plane at 5 GHz (Haynes, Caswell \& Simons 1979,
Altenhoff et al 1979). Later surveys, some with higher angular 
resolution at 1.4, 5  \& 10 GHz have detected this source 
(Handa et al 1987, Becker et
al 1994, Griffith et al 1995, Zoonematkermani et al 1990). 
It has also been mapped with high angular resolution
at 1.5, 4.9 \& 15 GHz (Garay et al 1993).

\subsubsection{Observational constraints}

 Among all available observational data for IRAS 18314--0720, those
most relevant for constraining the modelling are chosen 
based on their quality / sensitivity and the beam size (a
smaller beam size is preferred since we are dealing with
compact H II regions which are barely resolved in mid \&
far infrared wavebands). These include : IRAS PSC, LRS,
1.3 millimeter and the near infrared data for 
constructing the continuum SED. 
Observations at wavelengths longer than 1.3 mm 
have not been used to constrain the models because
at these wavelengths the free-free emission from
the gas will be a major contributor. Hence in order
to compare with the dust continuum emission predicted
by the models, one would need to subtract out the estimated 
free free emission from the measurements.
From the observations at 7 mm (Wood et al 1988) it is
clear that, the contamination due to 
the free free emission is negligible at 1.3 mm.
The infrared forbidden
line measurements of ions though detected in LRS, they
are only qualitative in nature, hence the only quantitative
data from the Kuiper Observatory has been used in the present study.
The distance to this source has been taken to be 9.4 kpc
from Chan, Henning \& Schreyer (1996). Accordingly the total luminosity
(from the observed SED) turns out to be $1.02 \times 10^{6}
L_{\odot}$. 
The radio continuum measurements at 5 GHz (VLA) and 10 GHz
(Nobeyama)
by Garay et al (1993)  and Handa et al (1987) respectively,
have been used for model fitting.

\subsubsection{Results of modelling}

   With the above observational constraints, the spherically
symmetric radiative transfer model D1 has been run (with
both DL \& MMP type of dust) exploring
various parameters described earlier. 

The resulting best fit model for the DL dust case, corresponds to : 
(i) a single ZAMS star of type O4 ($T_{eff}$ = 50,000 K)
as the embedded source;
(ii) an uniform density distribution (i.e. $n(r) = n_{0}$);
(iii) the radial optical depth at 100 $\mu$m, $\tau_{100}$ = 0.1; 
(iv) the gas to dust ratio by mass, 100~:~1;
(v) the density $n_{H} = 1.15 \times 10^{4} cm^{-3}$;
(vi) $R_{max}$ = 2.5 pc; (vii) $R_{min}$ = 0.05 pc; and
(viii) the dust composition, silicate : graphite : 
silicon carbide in 7.2 : 45.3 : 47.5 \% proportion.
This D1(DL) model predictions fit the observed SED extremely well,
which is shown in Figure 2. 
The predicted radio continuum emission at 5 GHz
(determined by G1(DL) run), using these parameters,
is only 324 mJy which is almost one tenth of the observed
value of 3.51 Jy. The increase in the gas to dust ratio
needed to bring the radio continuum emission close to the
observed value would be very unphysical.
The total cloud mass for this model turns out to be 
$1.85 \times 10^{4} M_{\odot}$ implying the $L/M$ ratio 
to be 55 $(L_{\odot}/M_{\odot})$.

On the other hand, the MMP dust case leads to the following
best fit parameters : 
(i) a single ZAMS star of type O4 ($T_{eff}$ = 50,000 K)
as the embedded source;
(ii) an uniform density distribution (i.e. $n(r) = n_{0}$);
(iii) the radial optical depth at 100 $\mu$m, $\tau_{100}$ = 0.1; 
(iv) the gas to dust ratio by mass, 300~:~1;
(v) the density $n_{H} = 3.76 \times 10^{4} cm^{-3}$;
(vi) $R_{max}$ = 2.5 pc; (vii) $R_{min}$ = 0.05 pc; and
(viii) the dust composition, silicate : graphite in  
11.5 : 88.5 \% proportion.
The fit to the observed SED for this  D1(MMP) model is also shown in 
Figure 2. Although the fit is reasonably acceptable, the
absorption feature at $\approx$ 10 $\mu$m predicted by this
model is much narrower than the LRS measurements.
However, the predictions for radio emission at 5 and 10 GHz 
in this case (G1(MMP)), viz., 3.56 Jy \& 4.56 Jy respectively,
match the observations very closely (3.51 Jy \& 4.51 Jy).
The total cloud mass in this case, turns out to be $5.78 \times 10^{4}
M_{\odot}$ implying a $L/M$ ratio of 17.3 $(L_{\odot}/M_{\odot})$.

 A comparison between the best fit parameters with DL \&
MMP dust, constrained by the observations of IRAS 18314--0720,
bring out
the following : most of the important parameters are identical
including even the radial optical depth. This is quite reassuring
to note that the geometry as well as the radial density 
distribution law are invariant to the type of dust
(among DL \& MMP) selected. Major differences exist for
the radio continuum emission predicted and the dust grain
composition. 
Considering the overall fit to the observed continuum
SED from the dust and the radio continuum emission from
the gas component, clearly the MMP is the favoured
self-consistent model for IRAS 18314-0720. Next, we consider
the additional information from the forbidden line emission
of ionized heavy elements.

  In order to predict the forbidden line emission from
various ionized species of the gas component in IRAS 18314--0720,
the G2 model calculations have been carried out.
Runs have been made with the best fit parameters
for both the models D1(DL) and D1(MMP).
The resulting line luminosities are presented in Table 1.
As described above (section 2.2.2) among the 27 relevant
lines, only those are included which have intensities
more that 1\% of the neighbouring continua (for
assumed instrumental spectral resolutions of
the ISO spectrometers).
The observed line strengths of IRAS 18314--0720, 
for the [S III] (18.7 $\mu$m),
[O III] (51.8 \& 88.4 $\mu$m), and [N III] (57.3 $\mu$m)
lines as reported by
Afflerbach, Churchwell \& Werner (1997; ACW), are
listed in Table 2. The corresponding model predictions 
(in identical units) for both G2(DL) and G2(MMP)
cases are presented in Table 2 for comparison. 
It is clear that both 
the models predict far less line emission (for all
the four lines) compared to the
observations. However, the DL case fares relatively
better than the MMP case. 
ACW have explained these line
emissions, originating from a region with electron density
$n_{e}$ = 825 $cm^{-3}$ and $T_{eff}$ = 35,000 K.
Their $n_{e}$ is too low compared to our
models. Perhaps that is the main reason for the discrepancy.
In order to verify that, another G2 run is carried out,
with the values of $n_{e}$, $T_{eff}$ and the elemental abundances
identical to those of ACW, but all other 
details identical to our D1(MMP) model. Results of this
run, G2(ACW), are also presented in Table 2.
The G2(ACW) predictions are very close to the observations
for the [S III] (18.7 $\mu$m) \& [O III] (88.4 $\mu$m)
lines and within a factor of two for the rest.

This confirms that a lower value of $n_{e}$ is mainly 
responsible for the higher observed line intensities
in general. This is not unexpected since the collisional
de-excitations will become important at higher
densities thereby reducing the probability of radiative decays.

The above implies that although an uniform density self-consistent
picture is able to explain the SED from the dust and the 
radio continuum emission from the gas, it fails to
explain details of fine structure line strengths for 
ionized heavy elements. Whereas the former suggests higher
densities, the latter favours a lower one. 
The detection of molecular maser sources  and CS line 
emission give additional
support to the existence of dense medium predicted by
our models. 
In actual source, the reality may lie somewhere in between,
viz., a mixture of dense clumps in a thinner inter-clump
medium.
We explore the role of clumpiness on resolving the issue
of fine structure line strengths. Consider the following 
simplistic scenario : clumps of constant density ($\rho_{1}$)
immersed in the inter-clump (lower density of $\rho_{2}$)
medium, with a volume filling factor of $f$. The clumps are
of uniform size and are uniformly distributed throughout
the interstellar cloud. This picture has three parameters,
viz., $\rho_{1}$, $\rho_{2}$  \& $f$.

In this approach, the inter-clump medium 
(with density $\rho_{2}$, as in the model of ACW)
will be mainly responsible for the fine structure line emission
(collisional deexcitation will be more important in the clumps).
On the other hand, the radio continuum emission will be dominated
by the region with higher $n_{e}^{2}$, i.e. the clumps.
In addition, if the fit to the observed continuum emission from
the dust (SED) is to remain intact, the effective total optical
depth (due to dust grains) has to match that from the 
radiative transfer model (DL or MMP).
Hence, there are three constraints to the scenario of clumpy
medium : $\rho_{2}$ dictated by the fine structure line data;
$< n_{e}^{2} >$ from the radio continuum; and the dust optical
depth from the continuum SED. 

Assuming both the clump and the inter-clump medium to be optically
thin in the radio; using the above three constraints; and DL type
dust; we obtain the following parameters
for IRAS 18314--0720. The clumps with a density of
1.34 $\times 10^{5}$ cm${}^{-3}$ are embedded in the inter-clump
medium (with a density of 825  cm${}^{-3}$; same as in ACW),
with a volume filling factor of 0.08. The detection of 98 GHz
line from the $CS$ molecule, whose excitation requires a critical
density $\approx$ 3--4$\times 10^5$ cm$^{-3}$, further supports the
above density of the clumps.

For the radiative transfer modelling of the continuum
SED to remain valid, the individual clumps must be optically
thin even at the lowest relevant wavelength (i.e. UV). This
condition translates the clump diameter to be less than
$10^{-3}$ parsec. For this size, the clumps
are found to be optically thin at radio wavelengths as well,
thus justifying our assumption of the same in calculating
the density ($\rho_{1}$) and the volume filling factor of the clumps.

 Thus, a self-consistent picture of IRAS 18314-0720
emerges with the DL model including clumpiness, which
explains all three major types of observational constraints,
viz., continuum SED, radio continuum \& fine structure line
emission from ionized heavy elements.

\subsection{IRAS 18355--0532}

 The IRAS PSC source 18355--0532
has flux densities of 24.6, 209, 1127 \& 1930 Jy
in 12, 25, 60 \& 100 $\mu$m bands respectively. This 
source is included in the original IRAS LRS Catalog.
An inspection of the LRS spectrum led to the identification 
/ possible identification of forbidden lines of neon \& sulphur ions 
(Jourdain de Muizon, Cox \& Lequeux 1990). 
More detailed analysis has identified and quantified
emission in a total of four lines from neon and sulphur
(Simpson \& Rubin 1990).
IRAS 18355--0532 corresponds to the 
RAFGL catalogue source no. 2211, which was
detected in 4.2, 11 \& 20 $\mu$m bands.
IRAS 18355--0532 was also included in the IRAS colour selected
sample of Chini et al (1986) for study at 1.3 mm continuum
and near infrared mapping (Chini, Krugel \& Wargau 1987).
Although the $CS$ emission at 98 GHz has been detected
from IRAS 18355--0532 (Bronfman, Nyman \& May 1996),
the searches for $H_{2}O$ (22.2 GHz) and methanol (6.6 GHz)
maser emission have been unsuccessful (Codella et al 1995; 
Schutte et al 1993).

The radio continuum emission associated with
IRAS 18355--0532 has been observed at various frequencies.
The IRAS PSC associates 18355--0532 with radio continuum sources 
of Parkes and Bonn
surveys of the Galactic plane at 5 GHz (Haynes, Caswell 
\& Simons 1979, Altenhoff et al 1979). Later surveys have also 
detected this source at 1.4, 5  \& 10 GHz 
(Handa et al 1987, Becker et
al 1994, Griffith et al 1995, Zoonematkermani et al 1990). 

\subsubsection{Observational constraints}

 The observations of IRAS 18355--0532 have been chosen to constrain
its modelling in the same fashion as in 
the case of IRAS 18314--0720.
The observed SED has been constructed by
the following measurements : IRAS PSC, IRAS LRS,
1.3 millimeter and the near infrared data.
The infrared forbidden line intensities have been taken from 
Simpson \& Rubin (1990), who have analysed and quantified 
the IRAS LRS measurements.
The distance adopted to this source has been taken to be 6.6 kpc
from Chini, Krugel \& Wargau (1987). 
The total luminosity of IRAS 18355--0532
from the observed SED is estimated to be $1.21 \times 10^{5}
L_{\odot}$. 
The radio continuum measurements at 5 GHz (VLA;
Becker et al 1994) and 10 GHz (Nobeyama; Handa et al 1987)
have been used for model fitting.

\subsubsection{Results of modelling}

The best fit parameters for the radiative transfer
modelling (D1(DL) case) of IRAS 18355--0532, are as follows : 
(i) a single ZAMS star of type O6.5 ($T_{eff}$ = 40,000 K)
as the embedded source;
(ii) an uniform density distribution (i.e. $n(r) = n_{0}$);
(iii) the radial optical depth at 100 $\mu$m, $\tau_{100}$ = 0.1; 
(iv) the gas to dust ratio by mass, 100~:~1;
(v) the density $n_{H} = 1.71 \times 10^{4} cm^{-3}$;
(vi) $R_{max}$ = 1.3 pc; (vii) $R_{min}$ = 0.007 pc; and
(viii) the dust composition, silicate : graphite : silicon carbide 
in 6.0 : 46.4 : 47.6 \% proportion.
The fit of this D1(DL) model to the observed SED is shown in
Figure 3. This fit is
very good for most of the spectral region, including the
10 $\mu$m feature.
Although, the model curve passes closely to the 25 $\mu$m 
point of the IRAS PSC, it deviates from 
the longer wavelength region ($\approx$ 15--22 $\mu$m)
of the LRS spectrum.
The radio continuum emission predicted by this model, at 5 GHz
is only 51 mJy which is again one tenth of the observed
value of 523 mJy.
Like we stated earlier (in case of IRAS 18314--0720), 
for IRAS 18355--0532 also,
we avoid increasing the gas to dust ratio
to bring the radio continuum emission closer to the
measurements, since that would require a very unphysical
value.
The total cloud mass for this model turns out to be 
$4.93 \times 10^{3} M_{\odot}$ implying the $L/M$ ratio 
to be 24.5 $(L_{\odot}/M_{\odot})$.

The D1(MMP) modelling leads to the following
best fit parameters for IRAS 18355--0532 : 
(i) a single ZAMS star of type O6.5 ($T_{eff}$ = 40,000 K)
as the embedded source;
(ii) an uniform density distribution (i.e. $n(r) = n_{0}$);
(iii) the radial optical depth at 100 $\mu$m, $\tau_{100}$ = 0.1; 
(iv) the gas to dust ratio by mass, 300~:~1;
(v) the density $n_{H} = 7.20 \times 10^{4} cm^{-3}$;
(vi) $R_{max}$ = 1.3pc ; (vii) $R_{min}$ = 0.018 pc; and
(viii) the dust composition, silicate : graphite 
in  12 : 88 \% proportion.
The fit to the observed SED for D1(MMP) model is also shown in 
Figure 3, which is reasonably good.
This model fits the longer
wavelength segment of the LRS spectrum quite well (which the
DL case could not), but the 10 $\mu$m
feature predicted is much narrower than the observed one. The
fit to 60 and 100 $\mu$m IRAS PSC points is slightly
poorer than the DL case.
This G1(MMP) case is much more successful than G1(DL), in
predicting the radio continuum emission. 
The predicted values for this case (G1(MMP)), 
are 872 mJy and 1.18 Jy at 5 and 10 GHz respectively, which are
very close to the observed numbers (523 mJy \& 1.61 Jy).
The mass of the cloud associated with IRAS 18355--0532,
for the parameters of this MMP model, is $1.56 \times 10^{4}
L_{\odot}$. This implies a $L/M$ ratio of 7.8 $(L_{\odot}/M_{\odot})$.

 Comparing the best fit parameters 
for IRAS 18355--0532, for the models 
with DL and MMP dust, one finds conclusions
similar to the earlier case of IRAS 18314--0720.
Most important parameters are identical, e.g.,
radial density distribution and the radial 
optical depth. The differences exist for
the predicted radio continuum and the dust grain
composition. 
 Once again, the MMP model is the favoured model for
IRAS 18355-0532, since it self consistently fits the
observed continuum SED as well as the radio continuum
emission, reasonably well.
Similar to the earlier source, here again we consider 
the data of fine structure lines of ionized heavy elements
in IRAS 18355-0532.

  The predictions from the G2 runs for IRAS 18355--0532,
corresponding to the above
mentioned best fit parameters of D1(DL) as well as 
D1(MMP) models are also presented in Table 1. 
Once again the line luminosities for only the detectable
infrared forbidden lines are included.
Simpson \& Rubin (1990; SR) have carefully analysed
the 8--22 $\mu$m IRAS LRS data for IRAS 18355--0532, as one
member of a large sample. They have
quantified the line intensities for [S IV] (10.5 $\mu$m),
[Ne II] (12.8 $\mu$m), [Ne III] (15.6 $\mu$m) and
[S III] (18.7 $\mu$m) lines. These are also presented in Table 3.
The corresponding model predictions 
(in identical units) for both G2(DL) and G2(MMP)
cases are listed in Table 3 for easy comparison. 
Both the models predict far less line emission (for all
the four lines) compared to the
observations. Just like in the earlier case of IRAS 18314--0720, 
for IRAS 18355--0532 also, the DL fares much better than the MMP. 

Since there are two pairs of lines from the same elements,
viz., S and Ne, the line ratios will be less sensitive to the
abundances. Whereas the measured
intensity ratios between [Ne II]/[Ne III] and
[S III]/[S IV] are 1.7 and 28 respectively, the same for
the G2(DL) model are 0.06 \& 3.1. It is interesting
to note that, if in the same G2(DL) model, the $T_{eff}$
is reduced to 28,000 K, then both the observed line ratios
are reproduced.

SR have modelled the line emissions from IRAS 18355--0532,
with electron density
$n_{e}$ = 3.16 $\times 10^{3}$ and $T_{eff}$ = 38,500 K.
This $n_{e}$ is very low compared to our models. 
In addition, their elemental abundances are different from ours.
In order to verify the hypothesis that,
the high value of $n_{e}$ is responsible for the
failure of our models to predict the line intensities,
another G2 run is carried out with the abundances,
$n_{e}$ and $T_{eff}$ values from SR, but all other details same as D1(MMP).
Predictions of this model, G2(SR), are also presented in Table 3. 
The G2(SR) predictions are reasonably close to the observations
for the [Ne III] (15.6 $\mu$m) \& [S III] (18.7 $\mu$m) 
lines but the other two line
intensities are down by a factor of $\approx $ 6.
This success of G2(SR) supports the above hypothesis 
about the value of $n_{e}$,
which is not unexpected since the collisional
de-excitations become less important at lower densities.

Once again, like in the case of IRAS 18314--0720,
for IRAS 18355--0532 also, an uniform density self-consistent
picture is able to explain the SED from dust and the 
radio continuum emission from the gas, but fails to
explain details of fine structure line strengths for 
ionized heavy elements. 
The detection of molecular maser sources  and CS line further
supports the existence of denser medium predicted by
our self-consistent models. 
Since again, a lower value of $n_{e}$ has been 
relatively more successful in predicting the forbidden line strengths,
we propose the possible scenario of clumpiness 
in IRAS 18355--0532 too, for 
resolution of the above problem, like the earlier case
of IRAS 18314--0720.

  Using an identical approach of incorporating clumpiness
in IRAS 18355-0532, as was used for the earlier source IRAS 
18314-0720, a physically meaningful solution has been
found corresponding to the DL scheme of modelling.
This solution corresponds to the following parameters :
$\rho_{1}$ = 2.12 $\times 10^{5}$ cm${}^{-3}$; 
$\rho_{2}$ = 3.16 $\times 10^{3}$ cm${}^{-3}$ 
(same as in SR) and $f$ = 0.067.
The detection of 98 GHz line from the $CS$ molecule 
further supports the above density inside the clumps.
Once again, arguing from the point of validity of the
DL modelling of the continuum SED, the upper limit on
the diameter of the clumps is set to 5.2 $\times 10^{-4}$
parsec. 

 Thus, even for IRAS 18355-0532, a self-consistent picture 
emerges with the DL model including clumpiness, which
explains all three major types of observational constraints.

\subsection{IRAS 18316--0602}

 The IRAS PSC source 18316--0602
has flux densities of 22.8, 138, 958 \& 2136 Jansky
in 12, 25, 60 \& 100 $\mu$m bands respectively. 
This source is included in the IRAS LRS Catalog. The
LRS spectrum shows the 10 $\mu$m silicate feature, but
no forbidden line or any feature due to 
the Polycyclic Aromatic Hydrocarbons
(Jourdain de Muizon, Cox \& Lequeux 1990). 
However, recent ISO-SWS measurements of IRAS 18316--0602
do show a very rich spectrum full of various solid state
molecular features (d'Hendecourt et al 1996; Dartois et al 1998).
IRAS PSC associates 18316--0602 with RAFGL 7009S, which 
was detected at 4.2, 11, 20 \& 27 $\mu$m.
Sub-millimeter and millimeter waveband continuum observations of 
IRAS 18316--0602 have been carried out at 450, 800, 850
and 1100 $\mu$m by Jenness, Scott \& Padman (1995) and 
McCutcheon et al (1995).
The 2.6 mm $CO$ line has also been detected from this source
by McCutcheon et al 1991.
The $CS$ and $NH_{3}$ emission have been detected from
from IRAS 18316--0602 (Bronfman et al 1996; Molinari et al 1996).
Searches for $H_{2}O$ and methanol maser emission from 
this source have also been successful 
(Brand et al 1994; van der Walt, Gaylard \& Macleod 1995;
Codella, Felli \& Natale 1996).

The radio continuum emission associated with
IRAS 18316--0602 has been observed 5 and 8 GHz
(Jenness, Scott \& Padman 1995; McCutcheon et al 1991; Kurtz,
Churchwell \& Wood 1994; Griffith et al 1995).

\subsubsection{Observational constraints}

 The observed SED for IRAS 18316--0602 has been generated from
the following : IRAS PSC, IRAS LRS, selected continua
from the ISO-SWS spectrum (3 \-- 8 $\mu$m) and all the
available sub-mm / mm observations (450 \-- 1100 $\mu$m). 
Radio continuum data at 5 and 8 GHz have been used as 
constraints for the modelling of this source.

The distance to this source has been taken to be 3.3 kpc
from Chan, Henning \& Schreyer (1996). 
The corresponding total luminosity
is $2.54 \times 10^{4} L_{\odot}$.

\subsubsection{Results of modelling}

The total luminosity of IRAS 18316--0602 corresponds 
to a single ZAMS star of type B0 ($T_{eff}$ = 30,900 K).
However, since a single ZAMS star as the embedded source leads 
to a poor fit to the observed SED as well as the 
radio continuum data, various clusters of ZAMS stars with a given
Initial Mass Function but a variable upper mass cut-off ($M_{u}$),
have been tried. After exploring the 
parameter space for D1(DL) modelling, the following
best fit parameters have been determined for 
IRAS 18355--0532 : 
(i) a cluster of ZAMS stars with an Initial Mass Function of the form
$N(M) \approx M^{-2.4}$, with the upper mass cut off, $M_{u}$, 
corresponding to the type B1.
(ii) an uniform density distribution (i.e. $n(r) = n_{0}$);
(iii) the radial optical depth at 100 $\mu$m, $\tau_{100}$ = 0.1; 
(iv) the gas to dust ratio by mass, 100~:~1;
(v) the density $n_{H} = 2.28 \times 10^{4} cm^{-3}$;
(vi) $R_{min}$ = 0.0001 pc; and
(vii) the dust composition, silicate : graphite : silicon
carbide in 71.8 : 28.2 : 0.0 \% proportion.
The above parameters are also consistent with the column density 
($N_{H_2}$) derived
from $CO$ measurements of McCutcheon et al 1991.
The fit of this D1(DL) model to the observed SED is shown in
Figure 4. This fit is
very good for most of the spectral region, particularly for
the longer wavelengths of the LRS spectrum.
The position of the predicted $\approx$ 10 $\mu$m feature 
is slightly to the shorter wavelength side compared to the
observations (LRS). Any attempt to ``align" this feature
towards the longer wavelength by increasing SiC dust
relative to the silicate dust, leads to very poor
fit to the 15--22 $\mu$m part of the spectrum (LRS data). 
Incidentally, the SiC dominant
dust composition corresponding to the best aligned feature,
is -- silicate : graphite : silicon carbide in 
0.9 : 11.0 : 88.1 \% proportion.
Although the predicted SED by the DL model passes acceptably close
to the ISO-SWS continuum at 4.0 $\mu$m,
the fit is rather poor at 5.0 \& 5.5 $\mu$m, the observed values being 
about 3 times the predictions.
The radio continuum emission predicted corresponding to
this model, at 5 GHz is about 2.16 mJy which is reasonably
close to the observed value of 2.7 mJy.
In fact a slight increase in the gas to dust ratio
can bring the model prediction exactly to the measured value.
The total cloud mass for this DL model turns out to be 
$3.88 \times 10^{3} M_{\odot}$ implying the $L/M$ ratio 
to be 6.55 $(L_{\odot}/M_{\odot})$.

The D1(MMP) modelling leads to the following
best fit parameters for IRAS 18316--0602 : 
(i) a cluster of ZAMS stars with an Initial Mass Function of the form
$N(M) \approx M^{-2.4}$, with the upper mass cut off, $M_{u}$, 
corresponding to the type B1.
(ii) an uniform density distribution (i.e. $n(r) = n_{0}$);
(iii) the radial optical depth at 100 $\mu$m, $\tau_{100}$ = 0.1; 
(iv) the gas to dust ratio by mass, 300~:~1;
(v) the density $n_{H} = 7.74 \times 10^{4} cm^{-3}$;
(vi) $R_{min}$ = 0.0005 pc; and
(vii) the dust composition, silicate : graphite 
in  11 : 89 \% proportion.
The fit to the observed SED of IRAS 18316--0602 by the D1(MMP) model 
is also shown in Figure 4. Although this fit crudely
represents the broad overall shape of the SED, it fails to reproduce
many details, particularly near the $\approx$ 10 $\mu$m
feature. The predicted feature in this MMP case is far too
narrow as well as shallow compared to the LRS data. At far infrared
wavelengths also (IRAS PSC 60 \& 100 $\mu$m) the predictions
are below the observations.
The predicted radio emission at 5 \& 8 GHz in this case (G1(MMP)) 
are 3.3  and 3.4 mJy respectively, which are in reasonable
agreement with the observations (2.7 \& 3.8 mJy; McCutcheon et al 
1991; Jenness, Scott \& Padman 1995).
The cloud mass for this MMP case is $1.31 \times 10^{4}
L_{\odot}$, leading to a value for the $L/M$ ratio
of 1.9 $(L_{\odot}/M_{\odot})$.

The D1(DL) is clearly the preferred model for IRAS 18316--0602.
The detection of maser sources associated with IRAS 18316--0602
is consistent with best fit model gas densities.

  The model G2 prediction (corresponding to the best fit
parameters of D1(DL) as well as 
D1(MMP) cases) for all the detectable
infrared forbidden line emission from IRAS 18316--0602
have been presented in Table 1, alongwith other two sources.
Although IRAS 18316--0602 has been studied
by the ISO-SWS spectrometer covering the entire
wavelength range of 2.5 -- 45 $\mu$m, the observations / data binning
have been carried out at a low resolution of 300 -- 500
(d'Hendecourt et al 1996; Dartois et al 1998). The few selected
narrow wavelength regions covered at higher resolution (1500
-- 2000; Dartois et al 1998) do not cover the lines
predicted to be detectable by our models (see Table 1).
In addition, the detectability criterion used by us uses
ISO-SWS observation modes with much higher resolution
(in the 12 -- 45 $\mu$m region) than these reported spectra
(i.e. even if the predicted lines were covered, they would
not have been detectable at this intermediate resolution
observational mode employed by Dartois et al 1998).

For IRAS 18316--0602 too,
like in the cases of IRAS 18314--0720 and
18355--0532, an uniform density self-consistent
picture is able to explain the SED from dust and the 
radio continuum emission from the gas. 
Unfortunately, no measurement exists to date, for
any infrared fine structure line for this source.
In case, IRAS 18316--0602 has been observed using the 
ISO-LWS spectrometer (in high resolution configuration), 
then our models can be qualified further.

\section{Conclusions}

   A simple yet self-consistent approach towards 
explaining observed spectral energy distribution of
interstellar clouds with embedded YSO's / compact H II
regions has been described. 
The radiation of the
embedded source/(s) is transported through the dust
and the gas components by different schemes in spherical
geometry. Two kinds of dust have been considered
(Draine \& Lee (DL); and Mezger, Mathis \&
Panagia (MMP)), each with its own variable composition.
Here, by self-consistent one means that the same geometric
and physical configuration fits the  observed
data for the emission from the dust (most of the
infrared, sub-mm, mm part of the SED) as well as the
emission from the gas (radio continuum).

  The effectiveness of this scheme has been assessed by
applying it to three
Galactic star forming regions associated with
IRAS 18314--0720, 18355--0532 and 18316--0602. They
cover a range of about 40 in luminosity of the
embedded source/(s). Relevant observational data
for these sources have been modelled to extract
information about their
physical size, density distribution
law, total optical depth and dust composition.
Interestingly, in all
these three cases, the best fit models correspond to
the uniform density distribution (for either DL or MMP dust). 
Similar conclusion about constant density envelopes,
has been drawn recently by Faison et al (1998)
for a sample of 10 Galactic compact H II regions.

For both IRAS 18314--0720 \& 18355--0532, the MMP dust
models are the favoured models, since they not only
give reasonably acceptable fits to the continuum SED,
but also explain the radio continuum data.

Even though SED \& radio continuum observations have been 
well explained by the above modelling, they predict much
lower intensities for fine structure lines of ionized
heavy elements, wherever measurements are available 
(IRAS 18314--0720 \& 18355--0532).
This discrepancy has been resolved by invoking clumpiness
in the interstellar medium. Two phase (clump / inter-clump)
models with DL type dust, have been successfully
constructed for IRAS 18314--0720 \& 18355--0532.

In the case of IRAS 18316--0602, DL is the preferred
model which gives a very good fit to the
observed SED, as well as predicts radio continuum 
emission which is consistent with the measurements.

\vskip 2cm
\centerline{\bf Acknowledgements}

 It is a pleasure to thank Gary Ferland for his help on several
occasions regarding the code CLOUDY;  and D. Narasimha for
clarifying certain doubts about radiative transfer. Members
of Infrared Astronomy Group are thanked for their comments.
The authors thank the anonymous referee for the comments
which improved the conclusions regarding clumpiness.

%
%

\newpage
\vskip 4cm
\centerline{\bf Appendix - 1}

\newcommand{\nc}{\newcommand}
\nc{\bec}{\begin{center}}
\nc{\enc}{\end{center}}
\nc{\beq}{\begin{equation}}
\nc{\enq}{\end{equation}}
\nc{\bei}{\begin{itemize}}
\nc{\eni}{\end{itemize}}
\nc{\bee}{\begin{enumerate}}
\nc{\ene}{\end{enumerate}}
\nc{\dep}{$\tau_{100\mu m}$}
\nc{\this}{4\pi r^{2}(4\pi\epsilon_{\nu}) n_e^{2}}
\nc{\that}{4\pi r^{2}\alpha n_e^{2}}

\def\iras{{\it IRAS{ }}}
\def\uchii{\hbox{UC{ }H{ }{\sevenrm II}{ }}}
\def\eg{{\it e.g.{ }},}
\def\ie{{\it i.e.{ }}}
\def\etc.{{\it etc.{ }}}
\def\etal{{\it et al.{ }}}
\def\arcmin{\hbox{$^\prime$}}
\def\arcsec{\hbox{$^{\prime\prime}$}}
\def\mic{$\mu$m}
\def\lsun{$L_\odot${}}
\def\rsun{$R_\odot${}}
\def\msun{$M_\odot${}}
\def\fir{far-infrared}
\def\nir{near-infrared}
\def\submm{submillimeter}

\def\boxit#1{\vbox{\hrule\hbox{\vrule\kern6pt
\vbox{\kern6pt#1\kern6pt}\kern6pt\vrule}\hrule}}

\vskip 1cm
{\bf A.1 Simple approach to radiative transfer through the gas (G1)}\\
\medskip
In this simple approach, the extent of the ionized region 
(in spherical geometry) is
determined by transporting the Lyman continuum photons 
(from the centrally located star / star cluster) through
the cloud including the effect of the dust, where they can exist
(as determined by their sublimation). 
The gas component of the cloud is assumed to consist of only hydrogen.
Next, the radio continuum
emission emerging from the cloud is calculated by transporting
the radio photons (free-free emission throughout the ionized medium 
of the cloud), 
through the entire cloud without making any approximation about the
optical thinness of the gas (i.e. self absorption is treated
appropriately).
The gas to dust coupling has been neglected.

\vskip 1cm
{\bf A.1.1 Extent of the ionized  region}\\
\medskip
The size of the H II region has been calculated by considering 
photoionization and recombination of hydrogen, 
along with the absorption due to the dust grains.
The presence of the dust reduces the size of the ionized
region,($R_{HII}$), compared to
that of pure gas Stromgren sphere considerably, 
depending on the density and the gas to dust ratio. 
The dust grains can exist in principle, only beyond a radial distance,
say $r_{subl}$,
depending on its sublimation temperature and the radiation
field due to the central source. In practice, the actual
distance beyond which the dust exists, say $r_{fit}$, is determined by 
the model fitting of the observed SED, by
radiative transport calculations through the dust (D1(DL) or D1(MMP)).
The $r_{fit}$ is often much larger than $r_{subl}$.

Hence, whether one encounters a dusty Stromgren sphere or not, 
is determined by the
type of the star / integrated spectrum of the cluster;
radial density distribution around the central star; and $r_{fit}$. 
We call it a Case A, if
the ionized region 
extends into the region where gas and dust coexist.
The other case of entire ionized region devoid of any dust grains
is termed Case B.
So for Case B, the extent of the H II  region can be obtained by solving the
equation,

\begin{equation}
-dN(r) = 4 \pi \alpha_{2} r^{2} n_{e}^{2}(r) dr
\end{equation}
where, $N$ is the number of Lyman continuum photons, $\alpha_{2}$ is
the recombination coefficient for hydrogen (for recombinations to all states
except the ground state) and $n_{e}$ is the number
density of electrons or $H^{+}$ ions (for a pure H II region),
 which in our case is the gas number density ($n_{g}$) and may be given by,

\begin{equation}
n_{g}~ =~ n_{0}^{'}(\frac{R_{min}}{r})^{m} , ~~~~~~m~=~0,1,2
\end{equation}

For m = 0, 1, 2  equation(1) can be solved easily by using the 
boundary condition

\begin{equation}
at~~~~ r~=~r_{*}, ~~~~~~~~~~~~N(0)~=~N_{Lyc}
\end{equation}

where, $N_{Lyc}$ is the total number of Lyman continuum photons emitted
per second by the embedded exciting star / star cluster
and $r_{*}$ is an effective stellar radius (with volume equal to the
sum of that of all the stars of the embedded cluster; as it turns out,
results are extremely insensitive to $r_{*}$).
\par
In case A however, the ionizing (Lyman continuum) photons experience further
attenuation due to direct absorption by the dust, 
so the modified radiation transfer
equation would be,
\begin{equation}
-dN(r)~=~4 \pi r^{2} \alpha_{2} n_{e}^{2}~dr~+~N(r) \tau_{Lyc}~dr
\end{equation}
where $\tau_{Lyc}$ refers to the optical depth of dust at $\lambda < 912 $\AA.
We solve the above equation, using the boundary conditions,
\begin{equation}
at~~~~ r~ = ~ R_{min}~~~~~~~~~~~~~N(R_{min})~=~N_{1}
\end{equation}
where $N_{1}$ is determined by using equation (1) and
\begin{equation}
at~~~~ r~ = ~ R_{HII}~~~~~~~~~~~~~N(R_{HII})~=~0
\end{equation}

\vskip 1cm
{\bf A.1.2 Calculation of continuum emission }\\
\medskip

With $R_{HII}$ properly determined, the radio continuum emission which occurs
due to the free-free emission  from the ions and electrons
can be calculated by using the formula (Spitzer 1978),
\begin{equation}
J_{\nu} = \int_{r_{*}}^{R_{HII}}\this e^{-\int_{r}^{R_{HII}}\kappa_{\nu}dr} dr
\end{equation}
where, the coefficients of emission $\epsilon_{\nu}$ and absorption
$\kappa_{\nu}$ are respectively given by,
\begin{equation}
\epsilon_{\nu}(erg/cm^{3}/sec/sr/Hz) = 5.44\times 10^{-39} g_{ff} Z_{i}^{2} n_{e
}n_{i}T^{-0.5}e^{-h\nu/kT}
\end{equation}
\begin{equation}
\kappa_{\nu}(1/cm) = 0.1731(1+0.130~log( T^{3/2}\nu^{-1}))Z_{i}^{2}n_{e}n_{i}T^{-3/2}
\nu^{-2}
\end{equation}
with, Gaunt factor ($g_{ff}$) given by,
\begin{equation}
g_{ff} = 9.77 (1+0.130~ log(~T^{3/2}\nu^{-1}))
\end{equation}

An electron temperature of 8000 K has been assumed for all calculations.
The radio continuum emission is computed at different frequencies
depending on the
availability of measurements for the particular astrophysical source
under study. The frequencies are typically between 5 and 10 GHz.
\par
%
%


\figcaption[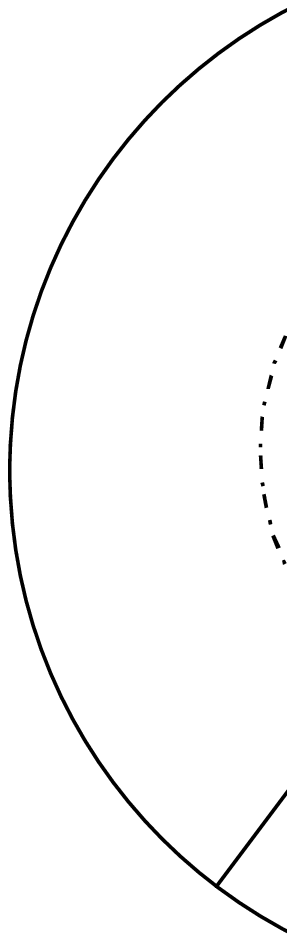]
{Schematic diagram of the model star forming region}

\figcaption[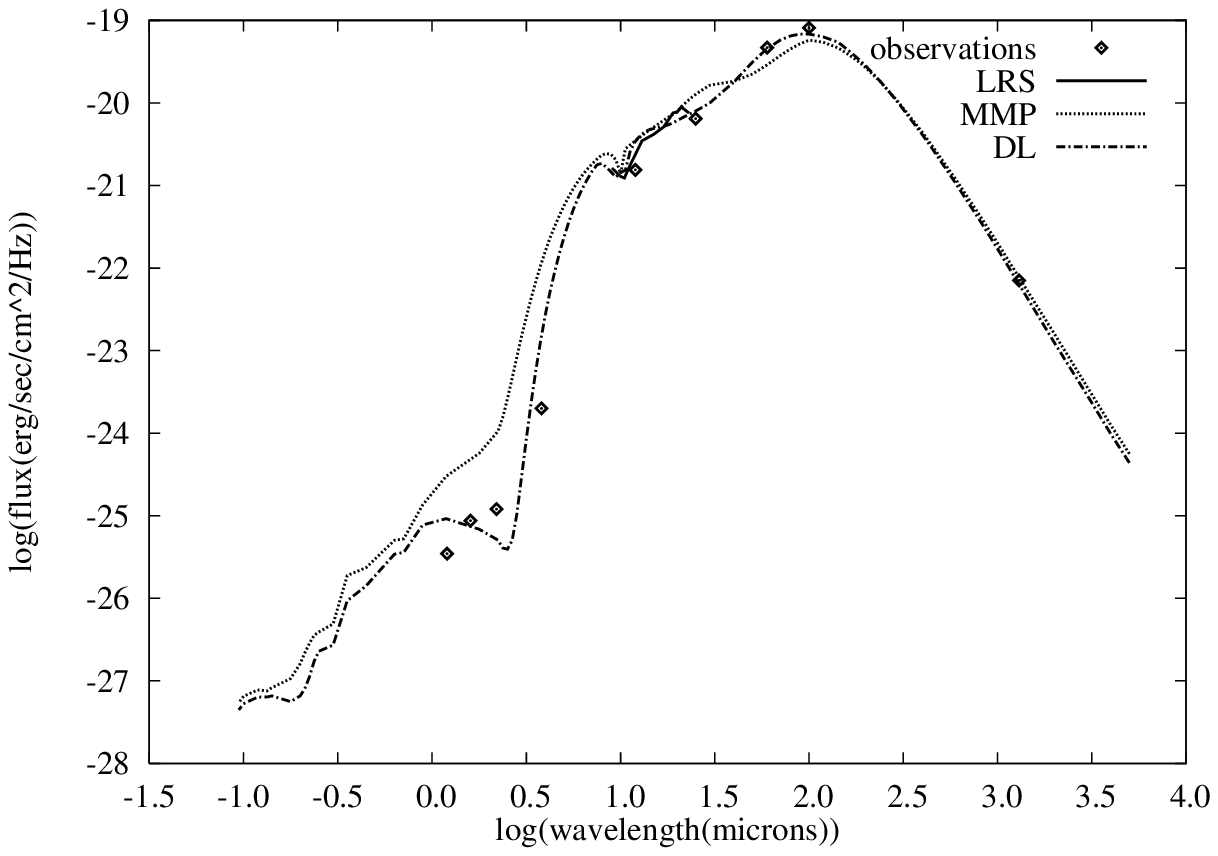]
{Spectral energy distribution for the source IRAS 18314-0720 ; the solid line represents LRS observations ; the dashed line represents model with MMP grains, the dash-dot-dash line represents model with DL grains and the symbol  $\Diamond$ represents other observations (see text). }

\figcaption[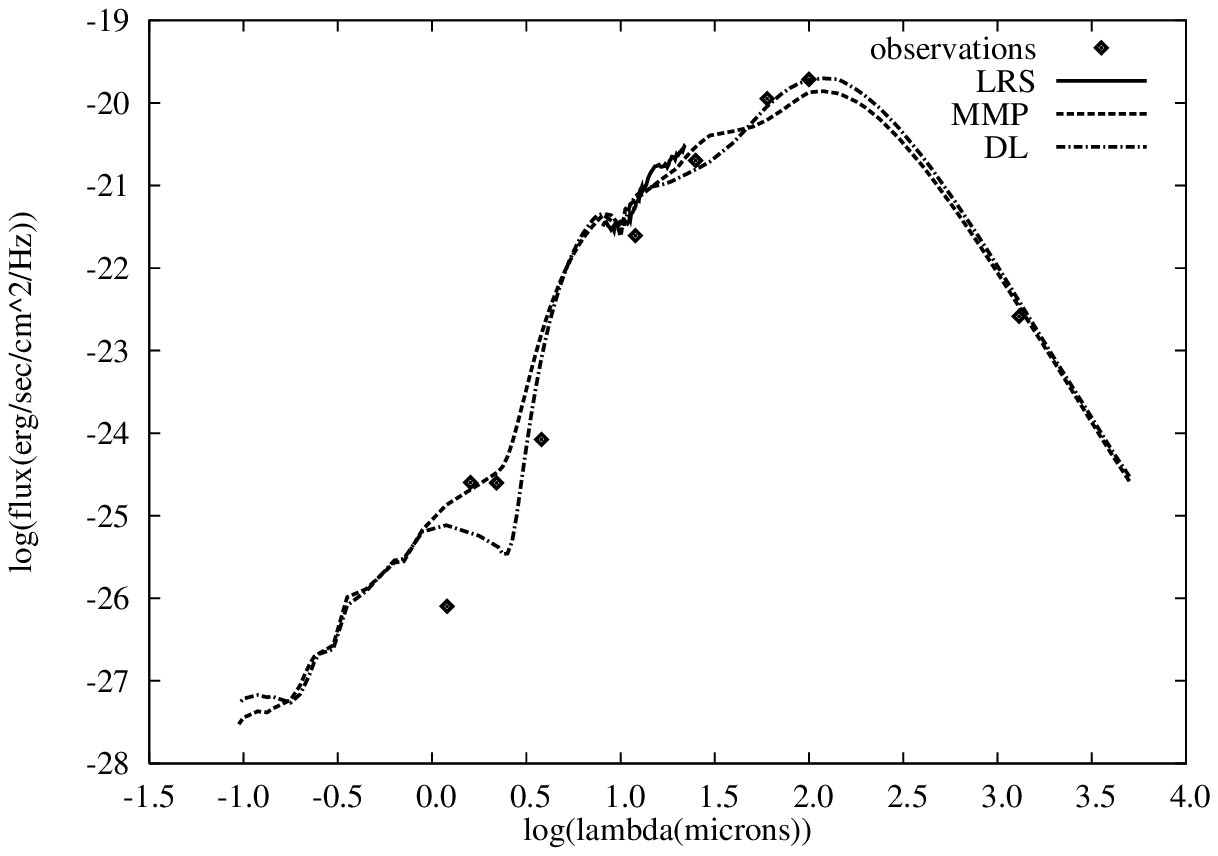]
{Spectral energy distribution for the source IRAS 18355-0532 ; the solid line represents LRS observations, the dashed line represents model with MMP grains, the dash-dot-dash line represents model with DL grains and the symbol  $\Diamond$ represents other observations (see text). }

\figcaption[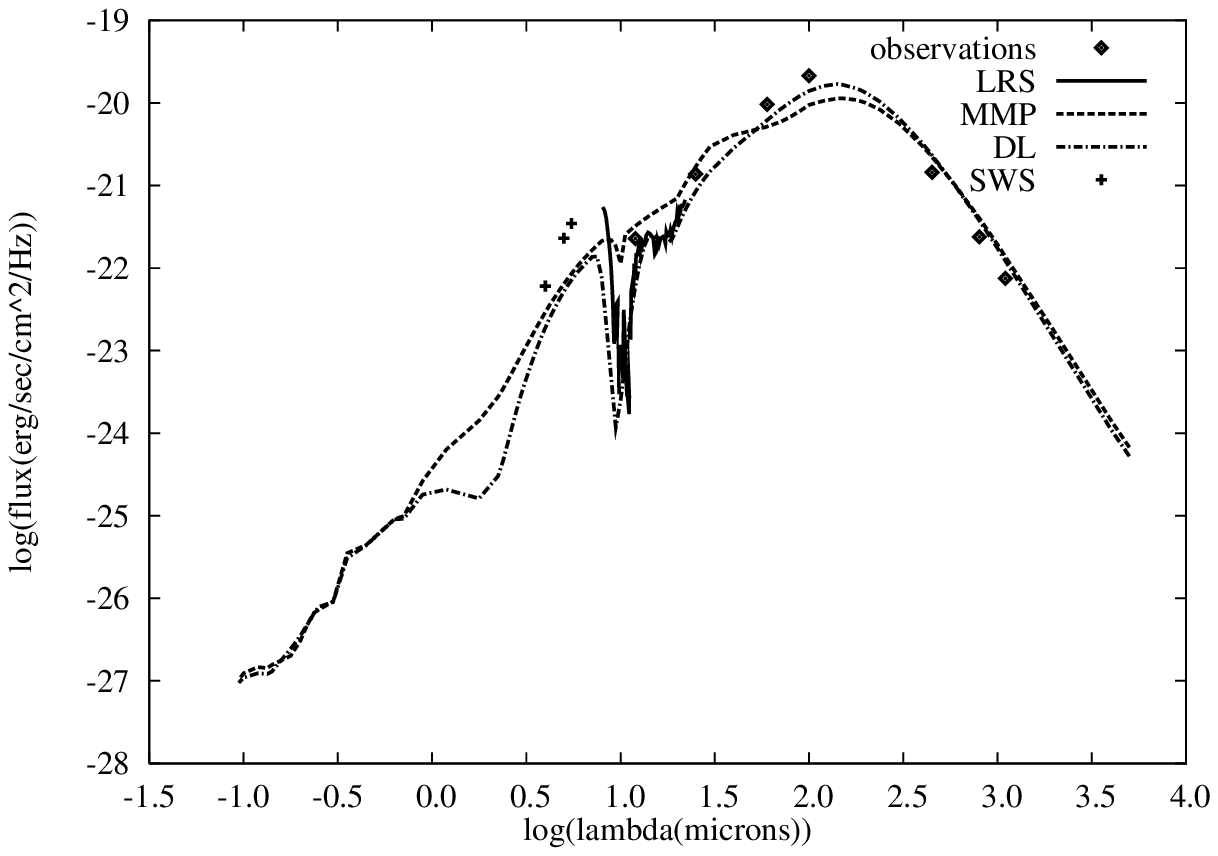]
{ Spectral energy distribution for the source IRAS 18316-0602 ; the solid line represents LRS observations, the dashed line represents model with MMP grains, the dash-dot-dash line represents model with DL grains, the symbol + represents SWS observations and the symbol  $\Diamond$ represents other observations (see text). }

\end{document}